\documentclass[12pt]{article}
\usepackage{epsfig}
\usepackage{rotating}
\begin{document}

\thispagestyle{empty}

\begin{center}
{\Large{\bfseries CAMAC subsystem and
user context utilities in ngdp framework
}}\\[10mm]
{\Large{\bfseries A.Yu.~Isupov}}\\[10mm]
{\itshape Veksler and Baldin Laboratory of High Energy Physics}\\[5mm]
{\itshape Joint Institute for Nuclear Research}
\end{center}


\newpage

\centerline{\bfseries Abstract}

\vspace*{5mm}

\noindent Isupov~A.Yu. \\
CAMAC subsystem and
user context utilities in ngdp framework

\vspace*{5mm}

The {\itshape ngdp} framework advanced topics are described.
Namely we consider
work with CAMAC hardware,
``selfflow'' nodes for the data acquisition systems with the
As--Soon--As--Possible policy, {\bfseries\itshape ng\_mm(4)} as
alternative to {\bfseries\itshape ng\_socket(4)}, the
control subsystem, user context utilities, events representation for the
ROOT package, test and debug nodes,
possible advancements for {\bfseries\itshape netgraph(4)}, etc.
It is shown that the {\itshape ngdp} is suitable for building lightweight
DAQ systems to handle CAMAC.

The investigation has been performed at the Veksler and Baldin Laboratory
of High Energy Physics, JINR.

\vspace*{1cm}

\centerline{\bfseries Аннотация}

\vspace*{5mm}

\noindent Исупов~А.Ю. \\
Работа с CAMAC и
утилиты режима задачи
в инфраструктурной системе ngdp

\vspace*{5mm}

Описаны продвинутые элементы
инфраструктурной системы (framework) {\itshape ngdp}:
работа с аппаратурой CAMAC, ``самотечные'' варианты
нод для систем сбора данных (DAQ) с соответствующей политикой, нода
{\bfseries\itshape ng\_mm(4)} как альтернатива сокету
{\bfseries\itshape ng\_socket(4)}, управляющая подсистема,
утилиты режима задачи, представление событий для пакета ROOT,
тестовые и отладочные ноды, возможные
дополнения собственно пакета {\bfseries\itshape netgraph(4)} и др.
Таким образом, {\itshape ngdp} пригодна для построения относительно простых
систем DAQ, обслуживающих CAMAC.

Работа выполнена в Лаборатории физики высоких энергий им. В.И.Векслера и
А.М.Балдина ОИЯИ.

\newpage



\setcounter{page}{1}

\section{Introduction}
\label{ngdp.intro}

\vspace*{-2mm}

\hspace*{4mm} The \cite{IsupJINRC10-34} paper describes only
basic design solutions and key elements of
the {\itshape ngdp} framework intended
for the data acquisition (DAQ) systems building. However many topics have not
been touched upon due to publishing limitations. So here we'll try to fill
this gap and consider the items, which allow to build some lightweight DAQ
systems to work with CAMAC using {\itshape ngdp} elements. Also we'll describe
some advanced topics like the control subsystem,
{\bfseries\itshape ng\_mm(4)} node as an alternative method to communicate
with {\bfseries\itshape netgraph(4)} graph in the kernel from the user context,
possible advancements for the {\bfseries\itshape netgraph(4)} itself,
test and debug nodes, etc.

Through the presented text
the
file and software package names are highlighted as
{\itshape italic text}, C and other languages constructions -- as
\verb|typewriter text|. Reference to the manual page named ``qwerty'' in the
9$^{\mbox{\small th}}$ section is printed as {\bfseries\itshape qwerty(9)},
reference to the sections in this paper -- as
``section \ref{ngdp.byprod.ng_fifos}''. Note also verbal
constructions like ``{\bfseries\itshape close(2)}d'' and
``\verb|mkpeer|ing'', which means ``closed by
{\bfseries\itshape close(2)}'' and ``peer making by \verb|mkpeer|''.
Subjects of substitution by actual values are
enclosed in the angle brackets: \verb|<setup>|, while optional
parameters are given in the square brackets: \verb|[mod [group]]|.
All mentioned trademarks are properties of their respective owners.

\vspace*{-5mm}

\section{Advanced topics implementation}
\label{ngdp.des_impl}



\vspace*{-3mm}

\subsection{{\bfseries\itshape netgraph(4)}
additionals, patches and improvements}
\label{ngdp.des_impl.intern}

\vspace*{-2mm}

\hspace*{4mm} The default maximal size of the {\bfseries\itshape netgraph(4)}
datagram can be too small to hold a packet with the event of some Mbytes.
Fortunately, all what we need
is changing of OS tunables (kernel and {\bfseries\itshape sysctl(8)} variables),
so recompile of the kernel is not required.

At the moment of writing this paper we saw the following
possible principial
advancements for the existing {\bfseries\itshape netgraph(4)} system:\\
$\bullet$ Generalization of the currently local scheme of the control
messages addressing
to be used remotely through the network. A special agent --
{\bfseries\itshape ng\_sv(4)}
(\underline{s}uper\underline{v}isor) node (see also
section~\ref{ngdp.des_impl.ctrl.sv}), -- is proposed to perform the
remote control messages delivering. The {\bfseries\itshape ng\_sv(4)} is
assumed to be launched on each involved computer. Also we should modify the
control messages delivery mechanism in the {\bfseries\itshape netgraph(4)}
code to change its behaviour to be some analogue of the ``default gateway''.
This means that all control messages with locally unknown (effectively
remote or wrong) addresses should be sent to the local
{\bfseries\itshape ng\_sv(4)} node for the decision on further delivering.
The remote delivery should use the connectionless (UDP) protocol, which
allows to communicate with an arbitrary number of hosts through the only one
{\bfseries\itshape ng\_ksocket(4)}.
We have also introduced
the \verb|NG_MKREMOTEMSG()| macro and patched the existing functions
\verb|ng_path_parse()|, \verb|ng_path2noderef()| and macro
\verb|NG_MKRESPONSE()|~.
Affected files: {\itshape ng\_base.c}~, {\itshape ng\_message.h}~.\\
$\bullet$ Possibility to insert some node between the two already connected
nodes. It is absent in the current {\bfseries\itshape netgraph(4)} system,
however it will be very useful for the
{\bfseries\itshape ng\_filter(4)} node implementation.
So we introduce the function\\
{\small \verb|int ng_insert(node_p, char *hook1, char *hook2, char *path1, char *path2)|~}.\\
Affected file: {\itshape ng\_base.c}~.

The following technical improvements in the form of
the corresponding patches are available:\\
1. {\bfseries\itshape ng\_socket(4)} is patched against the crash on the
netgraph datagram fragmented by TCP/IP. Affected file:
{\itshape ng\_socket.c}~.\\
2. {\bfseries\itshape ng\_ksocket(4)} is patched for \verb|NGM_KSOCKET_GETOPT|
control message proper working. Affected file: {\itshape ng\_ksocket.c}~.\\
3. The {\bfseries\itshape ngctl(8)} is
improved in some aspects:\\
 -- Buffer for \verb|write| command is increased to be large enough to hold the
increased {\bfseries\itshape netgraph(4)} datagram%
.\\
 -- Script variable \verb|$?| is introduced. It is substituted by a path
of the last node responding to {\bfseries\itshape ngctl(8)} control message.
\verb|$?| is very useful for \verb|name|ing of new
{\bfseries\itshape ng\_ksocket(4)} instance which appears after
the \verb|listen()|ing {\bfseries\itshape ng\_ksocket(4)} \verb|accept()|s.\\
 -- \verb|sleep| command is added to be used in scripts.\\
 -- A \verb|MsgCmd()| reaction on the \verb|EINPROGRESS| error from the
\verb|NgSendAsciiMsg()| is changed. Now \verb|MsgCmd()| waits for a response
during the timeout instead of the immediately error return. This behaviour
is needed to continue the script execution after
{\bfseries\itshape ng\_ksocket(4)} \verb|connect()|ing.\\
 -- New non--blocking function \verb|NgSendAsciiMsg_nb()| used in the
\verb|MsgCmd()| is introduced. It is able to wait for ascii--to--binary
conversion response up to timeout expiration instead of the infinite waiting in
\verb|NgSendAsciiMsg()| from {\itshape libnetgraph} library.
The remote delivering scheme of the control messages needs
this behaviour.\\
 -- The \verb|MsgRead()| function is patched to support remote control
messages delivering.\\
 -- \verb|msg| command now understands the remote path specification (see
section \ref{ngdp.des_impl.ctrl.sv}) in the form:
\verb|node_name@IP_address:hook1.hook2|. Note, that the so called generic
control messages are still only local as well as
the {\ttfamily mkpeer}, {\ttfamily shutdown}, {\ttfamily connect},
{\ttfamily rmhook}, {\ttfamily name}, {\ttfamily list}, {\ttfamily config},
{\ttfamily show}, {\ttfamily status}, and {\ttfamily types}
{\bfseries\itshape ngctl(8)} commands based on them.\\
 -- The \verb|dot| command implementation is refined in some aspects:
output of disconnected nodes is eliminated; fontsize is enlarged by 2~pt for
both the node and edge labels; explicit default fontname declaration
({\sffamily\bfseries Helvetica--Bold}) is added.\\
Affected files: {\itshape main.c}~, {\itshape msg.c}~,
{\itshape name.c}~, {\itshape write.c}~, {\itshape ngctl.h}~.\\
4. Ascii--to--binary and binary--to--ascii conversion schemes are revised
to handle the remote control messages delivering properly.
Affected file: {\itshape ng\_base.c}~.\\
5. The \verb|NgSendAsciiMsg()| and \verb|NgDeliverMsg()| functions from
the {\itshape libnetgraph} are patched to handle the remote control
messages delivering properly.
Affected file: {\itshape msg.c}~.

Let's note that during modifications of the basic
{\bfseries\itshape netgraph(4)} system we try to limit ourselves by only
absolutely necessary changes, as minimal as possible, in order to simplify
maintenance at the future {\bfseries\itshape netgraph(4)} version changes.

\vspace*{-5mm}

\subsection{{\bfseries\itshape ng\_mm(4)} node as an alternative for
{\bfseries\itshape ng\_socket(4)}}
\label{ngdp.des_impl.mm}

\vspace*{-2mm}

\hspace*{4mm} To allow a cheaper data injection into
{\bfseries\itshape netgraph(4)} system and their extraction from it
than it is possible to do by means of the standard method using the socket
mechanism provided by {\bfseries\itshape ng\_socket(4)} node,
we implement the {\bfseries\itshape ng\_mm(4)}
(for \underline{m}emory \underline{m}apping)
node type. It supplies two UNIX devices {\itshape /dev/mmr}\verb|<N>| and
{\itshape /dev/mmc}\verb|<N>|, whose instances with different unit numbers
\verb|<N>| and minors belong to different node type instances.
Both devices support {\bfseries\itshape open(2)}, {\bfseries\itshape close(2)},
{\bfseries\itshape ioctl(2)}, {\bfseries\itshape mmap(2)},
{\bfseries\itshape select(2)} / {\bfseries\itshape poll(2)}
system calls like usual UNIX devices do\footnote{
Note, we do not support {\bfseries\itshape read(2)} and
{\bfseries\itshape write(2)} because that contradicts to the main idea --
to eliminate the memory copying overhead as result of using
{\bfseries\itshape copyin(9)} / {\bfseries\itshape copyout(9)}.
}. When the user context process performs {\bfseries\itshape mmap(2)} of this
device it obtains a direct access to the packet
circle buffer allocated in the kernel memory by {\bfseries\itshape ng\_mm(4)}.
After that the process can read from the buffer of the so called ``raw''
device {\itshape /dev/mmr} and/or write to the buffer of the so called
``converted'' device {\itshape /dev/mmc}~. To synchronize the process with
the packets arrival in and departure from the buffers, the following
{\bfseries\itshape ioctl(2)} interface for these devices is provided
(first three commands are for a raw device while the last three ones -- for
the converted device):\\
\verb|MMGBUFRAW|, \verb|size_t| argument -- gets raw buffer size;\\
\verb|MMGROFFT|, \verb|size_t| argument -- gets offset of the next
packet ready to obtaining (process can be blocked up to the packet arrival);\\
\verb|MMRDDONE|, without argument -- says the packet reading done;\\
\verb|MMGBUFCONV|, \verb|size_t| argument -- gets the converted buffer size;\\
\verb|MMSSZGWOFFT|, \verb|size_t| argument -- sets the size of the packet to be
written and gets the offset where it should be written;\\
\verb|MMWRDONE|, without argument -- says the packet writing done.

So the data flow through the {\bfseries\itshape ng\_mm(4)} as follows: the
packets arrived on the \verb|in| hook are either placed into the raw buffer or
dropped, if the raw device was not opened; the process will be waken up,
if it was blocked on \verb|MMGROFFT| request; the oldest packet(s) will be
discarded, if the buffer lacks of space; however if the process
reads one of them, the newly arrived packet will be dropped. Immediately after
the process reports the end of the packet placement into the converted buffer,
this packet will be checked and, in case of approval, transmitted through the
\verb|out| hook. The packets arrived on the \verb|out| hook are always dropped.

To simplify \verb|mkpeer|ing in some situations, the
{\bfseries\itshape ng\_mm(4)} node supports the \verb|creat| hook, which can be
removed after \verb|in| or \verb|out| hook appearing, however \verb|in| or
\verb|out| hook can be used for \verb|mkpeer|ing, too.
If the process {\bfseries\itshape open(2)}s the raw device only and is sure
that it does not need the converted device (and vise versa, of course), then
for security reasons the process can ``catch'' the
converted device, so nobody (including this process itself) can open
it before the raw device is {\bfseries\itshape close(2)}d.
The raw device should be {\bfseries\itshape open(2)}ed with
\verb|O_EXCL| flag for catching the converted device (and vise versa).

Naturally, {\bfseries\itshape ngget(1)} and {\bfseries\itshape ngput(1)}
utilities (see section~\ref{ngdp.des_impl.utils}) with \verb|-A| and
\verb|-a| options use the {\bfseries\itshape ng\_mm(4)}'s interface.

The {\bfseries\itshape ng\_mm(4)} node supports a generic set of
{\bfseries\itshape netgraph(4)} control
messages as well as the following specific ones:\\
\verb|clristats| (\verb|clrostats|) -- clears the current statistics
  (numbers of \verb|data_packs|, \verb|data_bytes|, \verb|fails| and
  \verb|drops|) for the \verb|in|
  (\verb|out|) hook;\\
\verb|getistats| (\verb|getostats|) -- returns the current statistics
  for the \verb|in| (\verb|out|) hook;\\
\verb|getclristats| (\verb|getclrostats|) -- returns the current statistics
  and clears it for the \verb|in| (\verb|out|) hook;\\
\verb|getdev <struct ng_mm_getdev>| -- returns the raw and converted device
  names with unit numbers as C--strings \verb|<rawname>| and \verb|<convname>|.

\vspace*{-5mm}

\subsection{Nodes with ASAP policy}
\label{ngdp.des_impl.asap_nodes}

\vspace*{-2mm}

\hspace*{4mm} The As--Soon--As--Possible variants of the
{\bfseries\itshape ng\_fifo(4)},
{\bfseries\itshape ng\_em(4)} and {\bfseries\itshape ng\_pool(4)}
nodes \cite{IsupJINRC10-34} named {\bfseries\itshape ng\_fifos(4)},
{\bfseries\itshape ng\_ems(4)} ({\bfseries\itshape s} for \underline{s}imple) and
{\bfseries\itshape ng\_bp(4)} (for \underline{b}ranch \underline{p}oint
after {\itshape qdpb}'s terminology), are implemented to allow the
building of lightweight DAQ systems, for example \cite{Isup-rusNucl09lat},
\cite{IsupEXON09}. Generally speaking, ASAP (``selfflow'') policy means that
the data are processed immediately after obtaining and transmitted immediately
when ready. Therefore all the transfers are done
synchronously, if possible. Otherwise -- for example, if the \verb|rcvdata()|
of the destination node is locked, -- the {\bfseries\itshape netgraph(4)}
bufferizes the data intermediately and retransmits them later
invisibly for the nodes. So, a ``primary push'' from the data originator -- an
interrupt (IRQ) handler or some socket, -- is enough for a data packet to travel
through the whole DAQ system. The ``selfflow'' behaviour allows to avoid an
execution stream (kernel thread) in design of the ASAP client nodes
{\bfseries\itshape ng\_ems(4)},
{\bfseries\itshape ng\_bp(4)}%
.
In contrast, a complex DAQ systems with LAZY policy \cite{IsupJINRC10-34}
should have on each level at least
one execution stream able to send requests
to the bufferization server node {\bfseries\itshape ng\_fifo(4)} of the
data upstream level.

ASAP client nodes allow only one hook named
\verb|output| simultaneously, so to multiplicate
the output stream, we can use some
of {\bfseries\itshape ng\_tee(4)},
{\bfseries\itshape ng\_one2many(4)}, or {\bfseries\itshape ng\_fifos(4)}.
All ASAP nodes understand the generic {\bfseries\itshape netgraph(4)}
control messages.

\vspace*{-3mm}

\subsubsection{{\bfseries\itshape ng\_fifos(4)}: simple ``selfflow'' queue}
\label{ngdp.byprod.ng_fifos}

\vspace*{-2mm}

\hspace*{4mm}
The {\bfseries\itshape ng\_fifos(4)} node is able to:\\
$\bullet$ spawn \verb|listen()|ing \verb|ng_ksocket| at startup;\\
$\bullet$ spawn \verb|accept()|ing \verb|ng_ksocket|(s) at each connection
request from the known host(s) / port(s) up to the configured maximum, and/or\\
$\bullet$ accept hook connection from the local \verb|ng_socket|(s);\\
$\bullet$ emit each data packet obtained on the \verb|input| hook
as soon as possible
through all \verb|accept()|ing \verb|ng_ksocket|(s) and local
\verb|ng_socket|(s) currently connected;\\
$\bullet$ close \verb|accept()|ing \verb|ng_ksocket| at EOF notification
obtaining or connection loss.

The {\bfseries\itshape ng\_fifos(4)} node supports the same hooks and specific
set of control messages
as {\bfseries\itshape ng\_fifo(4)} \cite{IsupJINRC10-34} with exclusion
of \verb|setconf|, \verb|getconf|, \verb|outtype|, \verb|outpatt| coupled
with omitted debugging ability to self--generate packets.

\vspace*{-5mm}

\subsubsection{{\bfseries\itshape ng\_ems(4)}: event merger's
``selfflow'' implementation}
\label{ngdp.byprod.ng_ems}

\vspace*{-2mm}

\hspace*{4mm}
The {\bfseries\itshape ng\_fifos(4)} node
requires the corresponding clients, so
{\bfseries\itshape ng\_ems(4)} instead of requests issuing
is simply waiting for the data packets on the input channels.
When the arrived packets follow one of the configured merging rule(s), they
are merged%
. Such functionality does not require {\bfseries\itshape kthread(9)} usage,
however, it involves some complicated merging algorithm, too. Of course, id
marks \cite{IsupJINRC10-34} are supported by {\bfseries\itshape ng\_ems(4)},
because they are even more useful here than in the
{\bfseries\itshape ng\_em(4)} case, where, at least, the packet
numbers usually are guaranteed to be as required. Note, however, that
none of id marks is added by default in the node constructor, so at least
\verb|addid { name="num" }| control message should be sent to
{\bfseries\itshape ng\_ems(4)} for it works as expected.

Using the approach described for {\bfseries\itshape ng\_em(4)} \cite{IsupJINRC10-34}
we can compile the same {\bfseries\itshape ng\_ems(4)} source for both the
kernel and user contexts%
.
After strong debug sessions in the both contexts we are sure, that the
{\bfseries\itshape ng\_ems(4)} algorithm implementation is working now.

The {\bfseries\itshape ng\_ems(4)} node supports the same hooks and specific
set of control messages
as {\bfseries\itshape ng\_em(4)} excluding
\verb|settimo| and \verb|gettimo|.
The following control messages have slightly different meanings:\\
\verb|connect| -- checks the already supplied input channels and input type
  entries configuration, removes the
  unused input channels (if any) and connects the unconnected yet
  \verb|ng_defrag| subnodes to the upstream servers according to the current
  configuration;\\
\verb|start <int64_t outpacks>| -- starts packets accepting up to supplied
  \verb|<outpacks>| output packets
  will be merged (\verb|-1| means to accept infinitely);\\
\verb|stop| -- explicitly stops the packets accepting before the
  \verb|<outpacks>|
  are merged.

\vspace*{-5mm}

\subsubsection{{\bfseries\itshape ng\_bp(4)} node as ``selfflow'' version of
{\bfseries\itshape ng\_pool(4)}
}
\label{ngdp.byprod.ng_bp}

\vspace*{-2mm}

\hspace*{4mm} An ASAP version of {\bfseries\itshape ng\_pool(4)} node
\cite{IsupJINRC10-34} named {\bfseries\itshape ng\_bp(4)} is able to:\\
$\bullet$ launch the \verb|ng_defrag| subnode at each configured input channel,
this node in its turn launches the client \verb|ng_ksocket| node, which
\verb|connect()|s to the upstream ASAP server
corresponding to this channel;\\
$\bullet$ transmit all the packets, accepted in the input channel(s)
according to the configured rules, through the \verb|output| hook.

Each of the accepting rules is a \verb|struct tbl| and contains the input
packet type \verb|<in_type>|, which is allowed to arrive through any
configured input channel
with the corresponding set bit (equals to 1) in mask \verb|<mask>|.
The \verb|struct tbl| also contains the number \verb|last_num| of the last
obtained packet of type \verb|<in_type>|. A specific input
type value
\verb|-1| matches the arbitrary input type.
After {\bfseries\itshape ng\_bp(4)}'s \verb|mkpeer|ing the only one rule with
such wildcard input type and mask with all the bits set is defined. It is
equivalent to obtaining \verb|addcfg { in_type=-1 mask=-1 }| control message
at the early startup. Note, that the input type comparison is done in the same
order, in which the rules were added. So, the default wildcard rule will
always match first. Thus, to establish the rule(s) with the nonwildcard input
type, we should firstly remove the default rule -- f.e., by
\verb|delcfg { in_type=-1 }|. However, the last wildcard rule with some
restrictive mask, where only some bits are set, allows to
receipt packets of any type through only some input channels. The
default {\bfseries\itshape ng\_bp(4)}'s state after the fresh startup allows
the packets obtaining through any configured input channel%
.

The {\bfseries\itshape ng\_bp(4)} node supports the same hooks and specific
set of control messages
as {\bfseries\itshape ng\_pool(4)} excluding
\verb|settimo|, \verb|gettimo|.
The following control messages has slightly different meanings:\\
\verb|addcfg <struct tbl>| /
  \verb|delcfg <struct tbl>| -- adds / deletes the input packet type
  configuration entry;\\
\verb|getconf| -- returns the full current configuration of the input packet types;\\
\verb|connect| -- connects \verb|ng_defrag| subnodes to upstream servers
  according to the current configuration of the input channels;\\
\verb|start| -- allows to obtain packets through all the configured input
  channels;\\
\verb|stop| -- denies to obtain packets through all the configured input
  channels.

\vspace*{-3mm}

\subsection{Control subsystem}
\label{ngdp.des_impl.ctrl}

\vspace*{-2mm}

\hspace*{4mm} According to a big DAQ system scheme from \cite{IsupJINRC10-34},
the computers belonging to the SubEvB, EvB and pool levels
are controled from computers of the DAQ Operator group. The FEM level is
supervised from both the Slow Control group and FEM Control group. The
machines belonging to all the groups
(instead of levels) are autonomous. DAQ Operator group should be able
to propagate changes
in the {\itshape ngdp} system state very quickly. So the
corresponding software control subsystem, at least the slave side elements
on the very busy SubEvB, EvB and pool level machines, -- should be
implemented outside of preemptive scheduling. We have the
following options to do so:\\
$\bullet$ in the kernel context as some {\bfseries\itshape netgraph(4)}
node type: {\bfseries\itshape ng\_sv(4)} (for
\underline{s}uper\underline{v}isor);\\
$\bullet$ in the user context by programs with realtime priority for slave
elements and with any priorities for master elements on the usually idle
machines of the DAQ Operator group.

Note, that the user context realtime with the
guaranteed response time is practically impossible under usual UNIX--like
systems, so the last option can be unreachable.
The {\bfseries\itshape ng\_sv(4)} option is more attractive also for the
reasons of unification, startup schemes building and remote control
messages implementation (see section~\ref{ngdp.des_impl.intern}).

\vspace*{-5mm}

\subsubsection{{\bfseries\itshape ng\_sv(4)} prototype}
\label{ngdp.des_impl.ctrl.sv}

\vspace*{-2mm}

\hspace*{4mm} As a first step to {\bfseries\itshape ng\_sv(4)}
node with the functionality, described in
section~\ref{ngdp.des_impl.intern},
we implement some prototype, which is able to:\\[-8mm]
\begin{itemize}
\item spawn \verb|bind()|ed \verb|ng_ksocket| of UDP protocol at node
startup and respawn it after shutting down accidentally;\\[-8mm]
\item send and receive arbitrary remote control messages through this
\verb|ng_ksocket|;\\[-8mm]
\item send a \verb|test| remote control message to remote \verb|ng_sv| node
instance and respond to this message.\\[-8mm]
\end{itemize}

According to the control subsystem's requirements
the {\bfseries\itshape ng\_sv(4)} is ``permanent'' (survive
without hooks) node, around which a required {\itshape ngdp} graph could
be built during the DAQ startup or automatically recreated after troubles.

The prototype supports the following specific control messages:\\
\verb|bindaddr <struct sockaddr>| -- sets own IP address to perform
  \verb|bind()| in the same format as understood by the
  {\bfseries\itshape ng\_ksocket(4)} node;\\
\verb|testsend <char *ng_remote_path>| -- sends \verb|test| remote control
  message according to the supplied remote path specification
  \verb|<ng_remote_path>|.

The remote path specification extends the usual {\bfseries\itshape netgraph(4)}
absolute path specification in the following way: the nodename field before
``\verb|:|'' now can contain an IP address
after usual nodename itself and ``\verb|@|'' sign.
So, \verb|src2@192.168.10.15:out1| specification means the peer node of the
hook named \verb|out1|
of the node named \verb|src2| on the computer with IP address
\verb|192.168.10.15|.

\vspace*{-5mm}

\subsubsection{Control utilities}
\label{ngdp.des_impl.ctrl.util}

\vspace*{-2mm}

\hspace*{4mm} Some control utility could be implemented for each
introduced {\itshape ngdp} node type. The utility can have a graphical
user interface (GUI) to allow the end--user to be more comfortable than it is
possible by using the {\bfseries\itshape ngctl(8)} command string directly.
The control utility should be aware of the control messages specific to
the corresponding node type, as well as of the node type
defaults, parameter ranges, etc. Of course, the preferred
way to send control messages is to call {\bfseries\itshape ngctl(8)}
internally, however, the control utility could also use the
{\bfseries\itshape netgraph(3)} directly. The former approach similar to
the one used by the supervisor {\bfseries\itshape sv(1)} utility of the {\itshape qdpb}
system, which simply provides GUI over already existing command string tools.
There is a wide assortment of toolkits for the X Window System
\cite{X11ug}, \cite{X11nut}, which allow to implement any required GUI. Note,
that the control utilities implementation is reasonable at some mature stage
of the {\itshape ngdp} system, and it is beyond the scope of the present paper.

\vspace*{-5mm}

\subsection{User context utilities}
\label{ngdp.des_impl.utils}

\vspace*{-2mm}

\hspace*{4mm} As it was noted in \cite{IsupJINRC10-34},
any user context utilities previously
implemented for {\itshape qdpb} --
{\bfseries\itshape writer(1)} \cite{IsupJINRC01-116},
{\bfseries\itshape analyser(1)} \cite{IsupRNP05},
{\bfseries\itshape statman(1)}
\cite{IsupRNP03},
polarization calculators from polarimeter DAQ systems
\cite{IsupJINRC01-198}%
, \cite{Anis-PEChAJa04}, \cite{IsupCJP05}, -- are
still usable under {\itshape ngdp}, too, until
they are recompiled to be aware of the
redesigns mentioned in section~\ref{ngdp.servnodes.qdpb}. Here we consider
only utilities introduced
by {\itshape ngdp}.
We describe neither the command string options nor the reaction on signals
because the {\itshape ngdp} provides the manual pages for all the mentioned
utilities.
Note, that each utility exits 0 on success and $>0$ if an error occurs, and
has \verb|-h| flag means to write the usage to the standard error output
and exit successfully.

\vspace*{-5mm}

\subsubsection{{\bfseries\itshape ngget(1)}}
\label{ngdp.des_impl.utils.ngget}

\vspace*{-2mm}

\hspace*{4mm} The {\itshape ngget} is a utility for the packet stream
extraction from {\bfseries\itshape netgraph(4)}.

\vspace*{-2mm}


{\small
\begin{verbatim}
ngget [-f{<outfile>|-}] [-p{<pidfile>|-<template>XXXXX}] [-A|-a [-e]]
  [-l] [-d] [-r{<outrate>|-} [-v]] <peername> <peerhook> [<name> [<hook>]]
ngget -m [-f{<outfile>|-}] [-p{<pidfile>|-<template>XXXXX}] [-A|-a [-e]]
  [-l] [-d] [-r{<outrate>|-} [-v]] <peertype> <peerhook>
  [<peername> [<name> [<hook>]]]
\end{verbatim}
}

\vspace*{-3mm}


The {\itshape ngget} reads messages from the {\bfseries\itshape netgraph(4)}
data socket, optionally defragments
them into packets
and writes the packets to the
standard output. Thereof {\itshape ngget} is a service module to extract
the packet stream from the kernel graph into the user context.

In the first synopsis form the {\itshape ngget} connects the hook named
\verb|<hook>| (or ``\verb|in|'' if not supplied) of the newly created
{\bfseries\itshape ng\_socket(4)} named \verb|<name>|
(or ``\verb|ngget<PID>|'' if not supplied) to the hook \verb|<peerhook>| of
the already existing node \verb|<peername>|.

In the second synopsis form the {\itshape ngget} connects hook \verb|<hook>|
 of the newly created {\bfseries\itshape ng\_socket(4)} \verb|<name>|
to the hook \verb|<peerhook>| of the node with type \verb|<peertype>|,
newly created by the \verb|mkpeer| control message, and \verb|name|d
as \verb|<peername>|, if supplied.


\vspace*{-5mm}

\subsubsection{{\bfseries\itshape ngput(1)}}
\label{ngdp.des_impl.utils.ngput}

\vspace*{-2mm}

\hspace*{4mm} The {\itshape ngput} is a utility for the packet stream
injection to {\bfseries\itshape netgraph(4)}.

\vspace*{-2mm}


{\small
\begin{verbatim}
ngput [-l] [-d] [-c] [-p{<pidfile>|-<template>XXXXX}] [-A|-a [-e]]
   <peername> <peerhook> [<name> [<hook>]]
ngput -m [-l] [-d] [-c] [-p{<pidfile>|-<template>XXXXX}] [-A|-a [-e]]
   <peertype> <peerhook> [<peername> [<name> [<hook>]]]
\end{verbatim}
}

\vspace*{-5mm}


The {\itshape ngput} reads the packets
from the standard input and writes them to the {\bfseries\itshape netgraph(4)}
data socket. Thereof {\itshape ngput} is a service
module to inject the packet stream from the user context into the kernel graph.

In the first synopsis form the {\itshape ngput} connects hook \verb|<hook>|
(or ``\verb|out|'' if not supplied) of the newly created
{\bfseries\itshape ng\_socket(4)} \verb|<name>|
(or ``\verb|ngput<PID>|'' if not supplied) to the hook \verb|<peerhook>|
of the already existing node \verb|<peername>|.

In the second synopsis form the {\itshape ngput} connects hook \verb|<hook>|
of the newly created {\bfseries\itshape ng\_socket(4)} \verb|<name>|
to the hook \verb|<peerhook>| of the node with type \verb|<peertype>|,
newly created by the \verb|mkpeer| control message, and \verb|name|d
as \verb|<peername>|, if supplied.

\vspace*{-5mm}

\subsubsection{{\bfseries\itshape b2r(1)}
(\underline{B}inary--\underline{To}--\underline{R}OOT) converter}
\label{ngdp.b2r}

\vspace*{-2mm}

\hspace*{4mm}
The {\bfseries\itshape b2r(1)} reads the data packets from the standard input
and for each of them produces representation for the ROOT package \cite{ROOTproc}.
The {\bfseries\itshape b2r(1)} has three synopsis forms which correspond to the
following ROOT events' transfer variants:\\
$\bullet$ to remote client process(es) by ROOT \verb|TMessage| class through the
usual socket pair\footnote{
Using ROOT {\ttfamily TServerSocket} / {\ttfamily TSocket} wrappers.
};\\
$\bullet$ to the local child process by \verb|TBufferFile| through the
{\bfseries\itshape mmap(2)}ed memory or SysV IPC shared memory
mechanisms synchronized by the SysV IPC semaphores;\\
$\bullet$ to the local process by the data packets of the special type,
which encapsulate
\verb|fBuffer| of the \verb|TBufferFile| with the stored ROOT event,
through the standard output.

The {\bfseries\itshape b2r(1)}
avoids intermediate HDD storage, however, for flexibility reasons it is able
optionally to store ROOT events as
ROOT \verb|TTree| with single \verb|TBranch| into ROOT \verb|TFile|, too.
The {\itshape b2r.cxx}
source is written in terms of the only one ``high--level''
ROOT class to reach the event content independence
(see section~\ref{ngdp.ROOT_events} for details).
The object codes of the implementation as well as of ROOT
dictionary generated by {\bfseries\itshape rootcint(1)}
of this class and possibly of other involved classes,
should be linked (dynamically or statically) with the {\itshape b2r.o}.


\vspace*{-3mm}


{\small
\begin{verbatim}
b2r [-l] [-d] [-f{<outfile>|-}] [-s{<filesize>|-}] [-S{<splitlevel>|-}]
  [-p{<pidfile>|-<template>XXXXX}] [-a<addr>[ ...]] [-r{<outrate>|-} [-v]]
b2r -b<childname> -f{<outfile>|-} [-p{<pidfile>|-<template>XXXXX}]
  [-r{<outrate>|-} [-v]]
b2r -O [-m{<mmname>|-}] [-l] [-d] [-f{<outfile>|-}] [-r{<outrate>|-} [-v]]
  [-s{<filesize>|-}] [-S{<splitlevel>|-}] [-p{<pidfile>|-<template>XXXXX}]
\end{verbatim}
}

\vspace*{-4mm}


In the first synopsis form the {\itshape b2r}
sends the produced ROOT events to all the clients, which have already requested
registration on port 12340/tcp.
The number of simultaneously registered
clients is limited by the compiled--in value.

In the second synopsis form (so called batch or offline mode) the {\itshape b2r}
{\bfseries\itshape fork(2)}s and {\bfseries\itshape exec(3)}s
\verb|<childname>| child process (usually {\bfseries\itshape r2h(1)} in the
batch mode, too) with the same \verb|-f|, \verb|-r| and \verb|-v| options.
The \verb|-d|, \verb|-s|, and \verb|-S| options are ignored.

In the third synopsis form (so called output mode) the {\itshape b2r}
transfers the produced packets with ROOT events into the standard output.

\vspace*{-2mm}

\subsubsection{{\bfseries\itshape r2h(1)}
(\underline{R}OOT--\underline{To}--\underline{H}istogram(s)) converter}
\label{ngdp.r2h}

\vspace*{-2mm}

\hspace*{4mm} One of possible {\bfseries\itshape b2r(1)} clients is the
{\bfseries\itshape r2h(1)}, which fills some histograms from the
event--by--event data.
Like {\bfseries\itshape b2r(1)} the {\bfseries\itshape r2h(1)} is written
in terms of the only one ROOT class and obtains the event's data through the
interface provided by them
(see section~\ref{ngdp.ROOT_events} for details), so it is
event content independent, too. This means,
the {\itshape r2h.o} should be
linked in the same way as {\itshape b2r.o} (see section~\ref{ngdp.b2r}).
Also we implement the configuration and control protocol
{\bfseries\itshape r2h.conf(5)} for the runtime conversations between
{\bfseries\itshape r2h(1)} as server and its client(s). The
{\bfseries\itshape r2h(1)} startup configuration is performed from the file
written in terms of this protocol (see section~\ref{ngdp.r2hproto}).


\vspace*{-2mm}


{\small
\begin{verbatim}
r2h [-l] [-d] [-c{<cfgfile>|-}] [-f{<outfile>|-} [-s{<saverate>|-}]]
 [-p{<pidfile>|-<template>XXXXX}] [-r{<outrate>|-} [-v]] [-a<addr>[ ...]]
 [-A<addr>[ ...]] {[-I [-P]]|[<peerhost> [<peerport>]]}
r2h -b<shkey> -f{<outfile>|-} [-l] [-c{<cfgfile>|-}] [-a<addr>[ ...]]
 [-p{<pidfile>|-<template>XXXXX}] [-r{<outrate>|-} [-v]] [-A<addr>[ ...]]
\end{verbatim}
}

\vspace*{-2mm}


In the first synopsis form the {\itshape r2h} reads ROOT \verb|TMessage|s
from a server (f.e., {\bfseries\itshape b2r(1)})
through ROOT \verb|TSocket| connected to the port \verb|<peerport>| (12340 by
default) on the host \verb|<peerhost>|%
;
extracts event representation in the form of some compiled--in
ROOT class from each \verb|TMessage| obtained;
fills some histograms configured by \verb|<cfgfile>| file in the
{\bfseries\itshape r2h.conf(5)} format%
;
sends the requested histogram(s) to the corresponding registered client(s)
by \verb|TMessage|(s);
and optionally writes all the configured histograms to ROOT \verb|TFile|
\verb|<outfile>|. The {\itshape r2h} is \verb|listen()|ing on port 12341 for
the client registration requests. The number of simultaneously registered
clients is limited by the compiled--in value.

If the \verb|-I| option is specified, {\itshape r2h}
instead of \verb|TSocket| reads the standard input for the
data packets
of some compiled--in type, each of them should contain a serialized ROOT
event (f.e., produced by {\bfseries\itshape b2r(1)} in the output mode).
The {\itshape r2h} extracts this event using
the ROOT \verb|TBufferFile| with \verb|fBuffer|, which points to the packet's
body. With the \verb|-I| option the \verb|<peerhost>| and \verb|<peerport>|
arguments are ignored.

In the second synopsis form (so called batch or offline mode) the {\itshape r2h}
extracts each ready ROOT event using the ROOT \verb|TBufferFile| with
\verb|fBuffer|, which points to
the shared memory region with key \verb|<shkey>|, where the ROOT event was
stored by {\bfseries\itshape b2r(1)}. Other tasks are the
same as in the first synopsis form.
The memory synchronization is based on the SysV IPC semaphore mechanism.
The \verb|-d|,
\verb|-s| options and \verb|<peerhost>|, \verb|<peerport>| arguments are
ignored.

\vspace*{-5mm}

\subsubsection{
{\bfseries\itshape r2h.conf(5)}: control protocol for
{\bfseries\itshape r2h(1)}}
\label{ngdp.r2hproto}

\vspace*{-2mm}

\hspace*{4mm} The protocol for conversation between the histogram server
{\bfseries\itshape r2h(1)} and its
client(s) is \verb|#define|d in the {\itshape r2hproto.h} header
as follows:\\[1mm]
\begin{minipage}{70mm}
\begin{verbatim}
  #define	CMD_GET		10002
  #define	CMD_RESET	10003
  #define	CMD_DELETE	10004
  #define	CMD_BOOK1D	10005
  #define	CMD_BOOK2D	10006
  #define	CMD_RESETALL	10009
  #define	ACK_ERR		10010
  #define	ACK_OK		10011
\end{verbatim}
\end{minipage}\hfill\begin{minipage}{70mm}
\begin{verbatim}
  #define	CMD_QUIT	10012
  #define	CMD_LEAVE	10016
  #define	CMD_CONNECT	10017
  #define	ERR_NoSuchHist	1
  #define	ERR_UnknownCmd	2
  #define	ERR_PermDenied	3
  #define	ERR_ZipFailed	4
\end{verbatim}
\end{minipage}

\vspace*{1mm}

The \verb|What()| field of the ROOT \verb|TMessage| sent from the client to
{\bfseries\itshape r2h(1)} could contain any of \verb|CMD_*| values, while
the \verb|TMessage| body
should contain the command parameters in the C--string form. The
{\bfseries\itshape r2h(1)} can respond by:\\
\verb|ACK_ERR| with \verb|UInt_t error| (can have any of \verb|ERR_*|
values) in the \verb|TMessage| body;\\
\verb|kMESS_OBJECT| with \verb|TObject| of type \verb|TH1F| or
\verb|TH2F|\footnote{
Note, that it is a client's responsibility to determine an actual
{\ttfamily TObject}'s
type. For example, that can be done by comparison of {\ttfamily type\_info*}
of the obtained class ({\ttfamily TMessage::GetClass()->GetTypeInfo()}) with
the type of each {\ttfamily TObject} recognized by the client
({\ttfamily TObject::Class()->GetTypeInfo()}).
}; or\\
\verb|ACK_OK| without the body.

\noindent \verb|CMD_DELETE|, \verb|CMD_BOOK1D| and \verb|CMD_BOOK2D|
commands have parameter(s) as described below for their counterparts
\verb|Delete|, \verb|Book1d| and \verb|Book2d| in the
configuration language;\\
\verb|CMD_GET| and \verb|CMD_RESET| have a single histogram name
parameter \verb|hname|;\\
\verb|CMD_RESETALL|, \verb|CMD_QUIT| have no parameters;\\
\verb|CMD_LEAVE| (if implemented) and \verb|CMD_CONNECT| should be internal
for client;\\
\verb|Var|, \verb|Add2var| and \verb|Delvar| commands are currently internal
for server.

Clients from the hosts specified for {\bfseries\itshape r2h(1)} by \verb|-A|
option, have a permission to execute any \verb|CMD_*| commands mentioned
above (so called read--write access), while they from hosts specified by
\verb|-a| -- only \verb|CMD_GET| and \verb|CMD_QUIT| (so called read--only
access).

The configuration file in the {\bfseries\itshape r2h.conf(5)} format
consists of zero or more lines, delimited by the newline symbol. Lines could
be the comment lines, empty lines, and configuration lines, where the newline
symbol could be escaped by the backslash symbol to allow the line continuation.
The lines, which in the first position contain the ``\verb|#|'' (comment line)
and newline (empty line) symbols, are ignored.

Other lines should be configuration lines. Configuration lines concatenated
with all their continuations contain one or more fields, separated by space(s)
or tab(s). The fields themselves should not contain space(s) or tab(s).
The first field (\verb|command|) of the configuration line is a C--string and
represents the command name. The
second (if any) and following field(s) are command parameters.

Currently known commands are as follows:\\
{\ttfamily Var} --
declares the name \verb|varname| of the variable which could be histogrammed
in terms of the integer number triplet: channel \verb|chan|, module
\verb|mod| and group of modules \verb|group|. The event representation ROOT
class used by {\bfseries\itshape r2h(1)} should allow to obtain the variable's
value by such triplet (see section~\ref{ngdp.ROOT_events}).
The \verb|Var| has the following format%
:\\
\verb|Var varname chan [mod [group]]|~, for example:
\verb|Var adc0 2 0|~.\\
{\ttfamily Add2var} --
adds (yet another) triplet of channel \verb|chan|, module \verb|mod| and the
group of modules \verb|group| to the list of them belonging to the \verb|varname|
variable. So more than one detector channel could be united into the
same histogram.
The \verb|Add2var| has the following format:\\
\verb|Add2var varname chan [mod [group]]|~, for example:
\verb|Add2var adc0 3 1|~.\\
{\ttfamily Delvar} --
removes the variable named \verb|varname| and frees
all the corresponding memory. Its format is as follows:\\
\verb|Delvar varname|~, for example: \verb|Delvar adc0|~.\\
{\ttfamily Book1d} --
declares ROOT \verb|TH1F| histogram with name \verb|hname|,
id string \verb|fullhname|, variable to be histogrammed
\verb|varnameX|, number of bins \verb|chansX|, minimal \verb|minX| and
maximal \verb|maxX| bins. The histogram should be filled at each event arrival.
The \verb|Book1d| has the following format:\\
\verb|Book1d hname fullhname varnameX chansX minX maxX|~, for example:\\
\verb|Book1d adc0 ADC0 adc0 100 0 100|~.\\
{\ttfamily Book2d} --
declares ROOT \verb|TH2F| histogram with the name \verb|hname|,
id string \verb|fullhname|, variables to be histogrammed \verb|varnameX| and
\verb|varnameY|, number of bins \verb|chansX| and \verb|chansY|, minimal
\verb|minX|, \verb|minY| and maximal \verb|maxX|, \verb|maxY| bins. The histogram
should be filled at each event arrival.
The \verb|Book2d| has the following format:\\
{\small
\verb|Book2d hname fullhname varnameX chansX minX maxX varnameY chansY minY maxY|\\
}
for example: \verb|Book2d adc1_0 ADC1_0 adc1 200 0 200 adc0 300 0 300|~.\\
{\ttfamily Delete} --
removes the histogram named \verb|hname| and frees all the corresponding
memory. Its format is as follows:\\
\verb|Delete hname|~, for example: \verb|Delete adc1_0|~.

The configuration lines failed to parse or be processed are ignored while file
processing continues to the next line.

\vspace*{-5mm}

\subsubsection{{\bfseries\itshape histGUI(1)}:
standalone client for {\bfseries\itshape r2h(1)}}
\label{ngdp.histGUI}

\vspace*{-2mm}

\hspace*{4mm} One of possible {\bfseries\itshape r2h(1)} clients is
{\bfseries\itshape histGUI(1)} (for \underline{hist}ograms viewer with
\underline{GUI}), which requests histograms from the server and draws
them. The protocol for the client--server conversations described in
section~\ref{ngdp.r2hproto}
 allows {\bfseries\itshape histGUI(1)} to be independent on both
{\bfseries\itshape r2h(1)} internals and event representation ROOT class(es).
So {\bfseries\itshape histGUI(1)} depends on the standard ROOT classes only and
could be compiled under any OS equipped with the ROOT package libraries.

\vspace*{-3mm}


\begin{verbatim}
histGUI [-l] [-t<gui_sleep>] [<peerhost> [<peerport>]]
\end{verbatim}

\vspace*{-4mm}


The {\itshape histGUI} connects to
histogram server (f.e., {\bfseries\itshape r2h(1)}) on the port \verb|<peerport>|
(12341
by default) on the host \verb|<peerhost>|
through the ROOT \verb|TSocket|, launches ROOT \verb|TTimer| to do the required
redraws, displays the own GUI window and enters the ROOT eventloop.
After the user command obtaining (see section~\ref{ngdp.r2hproto}
for protocol description) the {\itshape histGUI} executes it or sends the
corresponding request to the server side, obtains the response, and draws the
histogram or reports the response error
status in the output viewer area.


{\itshape histGUI}'s GUI is designed to be self--explained. The main window
contains at least ``Exit'' button, the command input field (\verb|TGTextEntry|)
and output viewer area (\verb|TGTextView|). Single ROOT \verb|TCanvas| is
used to display all histograms after \verb|Get| command, while after
\verb|Getcont| each histogram is drawn in the own \verb|TCanvas| which
disappears after \verb|Stopall| or the corresponding \verb|Stop|.

\vspace*{-5mm}

\section{Scheme of event representation by ROOT class(es)}
\label{ngdp.ROOT_events}

\vspace*{-2mm}

\hspace*{4mm} Let us formulate the {\bfseries\itshape b2r(1)}
and {\bfseries\itshape r2h(1)} requirements to the interface to be provided
by the ROOT class (or number of classes) intended
to represent experimental events. Namely, the {\bfseries\itshape b2r(1)} needs
the following member functions:\\
$\bullet$ \verb|int pack2r(packet *)| fills the class data
members from the corresponding fields of the packet header and body, and
should be aware of all (possibly more than one) the involved packet types;\\
$\bullet$ \verb|void Clear()|\footnote{
It is assumed that its calling is cheaper than the full class instance
recreation by {\ttfamily delete} and {\ttfamily new}.
} resets the class instance into the unfilled state same as appeared after
constructor execution;\\
$\bullet$ \verb|uint16_t Get_type()| returns the value of \verb|type| data
member equal to the type of the packet, from which the last proper
filling of the class instance has been done;\\
$\bullet$ \verb|TObject *GetTObject(uint16_t t)| returns the pointer
(casted to \verb|TObject|\footnote{
The corresponding class should be derived from ROOT {\ttfamily TObject} base
class to allow such cast.
} po\-in\-ter to be used in \verb|TBufferFile::WriteObject()|) to the class
instance, which represents the event of supplied type \verb|t|;\\
$\bullet$ \verb|char *GetName()| returns the C--string representation
of the name of the class to be stored in ROOT \verb|TTree| for using by
\verb|TTree::Branch()|;\\
$\bullet$ \verb|void *GetAddr(void *addr)| returns the address of the pointer
to the class instance to be stored in ROOT \verb|TTree| for using by
\verb|TTree::Branch()|, where \verb|addr| is the pointer to instance of
the class known by {\bfseries\itshape b2r(1)};\\
$\bullet$ \verb|bool need_write()| returns the true value, if the class
instance is ready to be written into ROOT \verb|TFile|, and the false value
otherwise;\\
$\bullet$ (optionally) \verb|TObject *Get_sp()| returns the pointer
(casted to \verb|TObject*| to be used in \verb|TFile::WriteTObject()|)
to the class instance to be stored in ROOT \verb|TTree|.

The {\bfseries\itshape r2h(1)} needs the following:\\
$\bullet$ \verb|float Get_data(int chan, int mod, int group)| member function
returns the value to be histogrammed, which identified by the integer number
triplet: channel \verb|chan|, module \verb|mod|
and the group of modules \verb|group|;\\
$\bullet$ the ability to extract a class with this \verb|Get_data()| member
function from the obtained \verb|TMessage| or \verb|TBufferFile| class
instances in the unambiguous way.

The sources of both {\bfseries\itshape b2r(1)} and {\bfseries\itshape r2h(1)}
utilities are written to be event content independent, so each of them should
know about only one class, however, these classes
could be not necessarily the same.

In the simple cases, where we have the trigger data but not the data
coupled with the accelerator cycle, a single class could satisfy all the above
requirements,
f.e. {\ttfamily class Ekeyaf} \cite{Isup-rusNucl09lat}%
. Even if we have many trigger types, the single class scheme is still
suitable, if the trigger data contents do not differ essentially.
However, for the experimental setups on the cyclic accelerators
like Nuclotron (JINR, LHEP) we usually want to work with both: per event and
per burst information. For this complicated case we propose the following
approach.

The base class \verb|B<setup>| inherits from ROOT \verb|TObject| and
provides at least \verb|type|\footnote{
And, possibly, other data members are common for all event types, for example,
represent the packet header fields {\ttfamily num}, {\ttfamily len},
{\ttfamily sec}, {\ttfamily usec}, etc.
} data member and \verb|Get_type()|, \verb|Get_data()| member functions.
Each event type representation class -- f.e.,
\verb|E<setup>| (trigger event types), \verb|CB<setup>| (cycle begin event
type), \verb|CE<setup>| (cycle end event type) -- is derived from
\verb|B<setup>| and provides at least the specific data members of the type and
\verb|Get_data()| member function.
So \verb|B<setup>::Get_data()| simply calls its counterparts from the
event representer classes depending
on the data member \verb|B<setup>::type| value, while
the {\bfseries\itshape r2h(1)} simply does the so called upcast (transforms the
pointer to the obtained derived class into the pointer to \verb|B<setup>| base
class), so calls \verb|B<setup>::Get_data()|.

To produce ROOT \verb|TTree| with single \verb|TBranch|,
we provide \verb|S<setup>| ``container'' class derived from ROOT
\verb|TObject| to be a single ``leaf'' object per branch.
This class represents the whole accelerator cycle, thus, contains the following
data members: cycle number \verb|int cyc_num|;
\verb|TClonesArray *events| of trigger events and \verb|int Nevs| quantity of
them; and
\verb|CB<setup>|, \verb|CE<setup>| pointers -- as well as  the
member functions to store (\verb|int Fill(int cyc, int Nev)|,
\verb|void AddEv(E<setup> *)|, \verb|void AddCB(CB<setup> *)|,
~~~~~\verb|void AddCE(CE<setup> *)|) ~~~~~and ~~~~~retrieve\\
(\verb|int Get_cyc()|,
\verb|int Get_Nevs()|, \verb|CB<setup> *GetCB()|,
\verb|CE<setup> *GetCE()|, \verb|TClonesArray *Get_events()|) the corresponding data members.
The \verb|S<setup>| also provides \verb|void Clear()| member function.

After all we introduce the ``meta--class'' \verb|M<setup>| derived from ROOT
\verb|TObject|,
which contains
\verb|uint16_t type| of the event represented currently;
the pointer to container \verb|S<setup> *sp|;
the instance of each known event representer
\verb|E<setup> ev|, \verb|CB<setup> cb|, \verb|CE<setup> ce|;
and provides the full interface required by {\bfseries\itshape b2r(1)}.
Na\-me\-ly, \verb|Get_type()| returns \verb|M<setup>::type| value;
\verb|GetTObject(t)| -- one of \verb|&ev|,
\verb|&cb|, \verb|&ce| depending on \verb|t| supplied;
\verb|GetName()| returns name of \verb|S<setup>|;
\verb|Get_sp()| and \verb|GetAddr()|
return \verb|sp| and \verb|&sp| with the required casts.

The most sophisticated job
is done by \verb|pack2r()| called once per each data packet arrival.
The \verb|pack2r()| assumes the data for each accelerator cycle are
starting from the cycle begin packet (if any), contain zero or more trigger
packet(s), and terminated by the cycle end packet.
After each \verb|pack2r()| return the event representer, corresponding to the
type of the arrived packet, is ready to be sent to the
{\bfseries\itshape b2r(1)} registered client(s). After the cycle end packet
processing by \verb|pack2r()|, the \verb|S<setup>| is ready to be stored
into ROOT \verb|TFile| and \verb|need_write()| returns the true value, so
{\bfseries\itshape b2r(1)} writes ROOT \verb|TFile| only if this condition
is satisfied.

\vspace*{-5mm}

\section{Work with CAMAC hardware}


\label{ngdp.camac_package}

\vspace*{-2mm}

\hspace*{4mm} As it was noted in \cite{IsupJINRC01-116}, the {\itshape qdpb}
framework uses the {\itshape camac} package \cite{GrOllat} as the CAMAC subsystem
implementation. The {\itshape ngdp} inherits this approach, so lets briefly
explain the current design features of the {\itshape camac} package according
to its manual pages.
The code, which handles a specific CAMAC crate
controller / computer interface card pair, is separated from the so called
CAMAC facility {\bfseries\itshape camac(4)} in the OS kernel. This design
allows to add support for an arbitrary new CAMAC controller/adapter pair
very easily. The {\bfseries\itshape camac(4)} supports the
interface to software objects of two kinds, mentioned below as
``CAMAC drivers'' and ``CAMAC modules''. The
CAMAC drivers (see {\bfseries\itshape camacdrv(4)})
are kernel device drivers in the KLD module form,
and work with some specific CAMAC hardware like CAMAC crate controllers.
CAMAC modules (see {\bfseries\itshape camacmod(9)}) are intended to handle
interrupts from CAMAC hardware, and usually have a KLD module form, too
(see below for details). The {\bfseries\itshape camac(4)} realizes some
abstraction level for CAMAC operations and therefore hides the CAMAC hardware
specifics from the user and
kernel contexts. Each CAMAC operation is performed at the CAMAC
address, which uniquely represents the hardware corresponder of the operation.
Drivers
should be registered
in the
{\bfseries\itshape camac(4)} before using. Driver registration consists of
the branch number assignment to the named driver. The
CAMAC driver can implement up to 4 methods:
\verb|subr()|, \verb|setup()|, \verb|conf()|, and
\verb|test()| (former is mandatory while others are optional).
The \verb|subr()| function implements the CAMAC operations themselves, f.e.
CCCZ -- generates the dataway initialize (Z), CCCC -- generates the crate
clear (C), CFSA -- performs a single CAMAC action, etc., and can be
achieved from both contexts by the corresponding interfaces.
The \verb|setup()| function attaches the CAMAC module of the user provided
interrupt handler to the CAMAC driver.
The CAMAC module should be registered in the
{\bfseries\itshape camac(4)} before using and deregistered before
unloading from the kernel. The CAMAC module can implement up to 4 methods:
\verb|hand()|, \verb|oper()|, \verb|conf()|, and
\verb|test()| (former is mandatory while others are optional).
The \verb|hand()| function is an interrupt handler itself and should address
only one CAMAC driver, specified during the module configuration.
The \verb|oper()| function is a
user interface to the module and can
access CAMAC. The CAMAC module should also meet the requirements of
{\bfseries\itshape ng\_camacsrc(4)} (see section~\ref{ngdp.camacsrc}) to be
able to inject the data into {\itshape ngdp} graph. Typically the CAMAC
module sources are split to the number of files to isolate the
hardware dependent code parts. Namely, we define some macros --
to initialize CAMAC,
to recognize the
trigger type, to perform read/reset CAMAC for each trigger type, etc.,
so \verb|hand()| function can be written in the hardware independent manner.
This macro set can be written ``manually'' for compact stable setups, or
generated by means of configurable representation of the CAMAC
hardware \cite{IsupJINRC01-199} for more extensive
and/or changed from the run to run spectrometers.

\vspace*{-5mm}

\subsection{
{\bfseries\itshape ng\_camacsrc(4)}: {\itshape ngdp} CAMAC interface}
\label{ngdp.camacsrc}

\vspace*{-2mm}

\hspace*{4mm} The {\bfseries\itshape ng\_camacsrc(4)} node type allows to
inject data packets from a CAMAC interrupt handler into
{\bfseries\itshape netgraph(4)} as data messages.
This means that we are able to use {\itshape ngdp} for building
DAQ systems, which work with CAMAC, as well as to do some sophisticated
testing of {\itshape ngdp} itself. This node supports the single hook
\verb|output|, which can be connected to, for example, the \verb|input| hook
of the {\bfseries\itshape ng\_fifos(4)} node (see
section~\ref{ngdp.byprod.ng_fifos}). The {\bfseries\itshape ng\_camacsrc(4)}
sends the data packets
in the interrupt handler context, where furter data processing is undesirable.
So the data should always be queued by {\bfseries\itshape netgraph(4)}
for later delivering to decouple from this context. Since by default the
{\bfseries\itshape netgraph(4)} delivers the data synchronously, we should
set \verb|HK_QUEUE| flag on the peer hook of the
{\bfseries\itshape ng\_camacsrc(4)} \verb|output|
hook during the hooks connection.

The {\bfseries\itshape ng\_camacsrc(4)} provides the kernel--wide interface
function%
\footnote{
Prototyped as {\ttfamily int ng\_camacsrc\_send(camacsrc\_p msp)}~.
} to send the data packet through the
\verb|output| hook, which returns 0 on success and error code otherwise.
The function argument (\verb|msp|) is the pointer to the instance of the node's
private structure which belongs to the node instance, through whose
\verb|output| hook the interrupt handler wants to send the data.
The {\bfseries\itshape ng\_camacsrc(4)} node declares its own version number
in the {\itshape ng\_camacsrc\_ver.h} to allow the dependent KLD modules to
declare such dependence and
resolve \verb|ng_camacsrc_send| symbol during the linkage into the kernel.
In contrast, the {\bfseries\itshape ng\_camacsrc(4)} calls
functions, provided by an interrupt handler, indirectly (by function
pointers) to avoid cross--dependencies.

During
processing of the \verb|connect| control message
(see below) the {\bfseries\itshape ng\_camacsrc(4)} obtains the
corresponding function pointers\footnote{
These pointers are stored in the
specific data structures of the CAMAC driver and the
corresponding CAMAC module and can be found by known CAMAC driver name using
the {\bfseries\itshape camac(4)} interface.
} and calls the interrupt handler's
\verb|oper()| function with the following subfunction codes:\\
\verb|NG_CAMACSRC_SETMSP| -- provides \verb|msp| address for the interrupt
handler, and\\
\verb|NG_CAMACSRC_GETBUF| -- obtains the address of the ready packet buffer
from it.

At receiving the \verb|disconn| control message (see below) and
during the shutdown sequence the {\bfseries\itshape ng\_camacsrc(4)} calls
\verb|oper()| with \verb|NG_CAMACSRC_CLOSE| subfunction code to notify
the interrupt handler about its own retirement.

So, the interrupt handler named \verb|<camac_module>| should provide these
three subfunction entries in its own \verb|oper()|
function realization. After the existing CAMAC interrupt handler
implementations
used in {\itshape qdpb} based DAQ systems it is also assumed, that the
data produced in the packet form\footnote{
It will be very easy to move the packet encapsulation by
{\ttfamily make\_pack()} function (see {\bfseries\itshape packet(9)}) into
the generic {\bfseries\itshape ng\_camacsrc(4)} code.
} -- only to be able to get the data length from the known place in the
data packet.

The described scheme allows to use
more than one instance of {\bfseries\itshape ng\_camacsrc(4)} nodes on
the same computer, if we have more than one CAMAC driver and the corresponding
CAMAC interrupt handler\footnote{
Possibly we can generalize this model to use more than one
{\bfseries\itshape ng\_camacsrc(4)} instance with the same
interrupt handler to separate the packet streams on the early stage or
to enlarge the overall throughput.
}.

In the {\itshape qdpb} system the control over \verb|<camac_module>| CAMAC
interrupt handler was done only by locally executed utility
\verb|<camac_module>|{\bfseries\itshape oper(8)}, which
calls \verb|<camac_module>_oper()| with some required subfunction codes.
This code set can be divided to ``generic'' (supported by approximately
each CAMAC module) and ``specific'' (all others) parts. At least the following
subfunction codes can be considered as ``generic'': \verb|INIT|, \verb|FINISH|,
\verb|START|, \verb|STOP|, \verb|CNTCL|, \verb|QUECL|. To allow cheaper
and more flexible (in particular, remote) control over the CAMAC
interrupt handler, the {\bfseries\itshape ng\_camacsrc(4)} understands a number
of control messages (see below), whose arrival leads to performing each
``generic'' subfunction and even more. Note the same subfunctions can be
easily implemented by the control packet mechanism, too.

The {\bfseries\itshape ng\_camacsrc(4)} node supports the following specific
control messages:\\
\verb|getclrstats| -- returns the current statistics (numbers of \verb|packets_out|,
  \verb|bytes_out|, and \verb|fails|) and
  clears it;\\
\verb|getstats| / \verb|clrstats| -- returns / clears the current statistics
  (the same values);\\
\verb|connect <char *camac_driver_name>| -- establishes a link
  with the CAMAC interrupt handler hosted on the CAMAC driver with name
  \verb|<camac_driver_name>|%
;\\
\verb|disconn| -- breaks the link established by the \verb|connect| control
  message;\\
\verb|init|, \verb|finish|, \verb|start|, \verb|stop|, \verb|cntcl|,
  \verb|quecl| -- performs \verb|<camac_module>_oper()| call with the
  corresponding subfunction codes, which lead to: CAMAC hardware
  initialization (\verb|INIT|), CAMAC hardware preparation to power off
  (\verb|FINISH|), data acquisition start (\verb|START|) and stop
  (\verb|STOP|), clean of the interrupt handler internal counters
  (\verb|CNTCL|) and event buffers (\verb|QUECL|);\\
\verb|getmodstat| / \verb|getmodconf| -- returns some generic part of the
  internal statistic / configuration
  structure of the interrupt handler (\verb|GETPSTAT| / \verb|GETPCONF|);\\

\vspace*{-11mm}

\section{Conclusions}
\label{ngdp.5}

\vspace*{-3mm}

\hspace*{4mm} 
The {\itshape ngdp} framework allows to build full--sized modular
distributed DAQ systems in a very efficient way with the minimal design and
maintenance efforts due to the already
implemented interface to CAMAC as well as
binary--to--ROOT and ROOT--to--histogram conversion utilities. The proposed
representation scheme of the experimental events for the ROOT package
allows to handle many trigger types and data coupled with the accelerator
cycle, and hides this complexity from the mentioned converters.

\vspace*{-5mm}

\section*{Acknowledgements}
\label{ngdp.Ack}

\vspace*{-3mm}

\hspace*{4mm} The author has a pleasure to thank K.I.Gritsaj for useful
discussions and many years of {\itshape camac} package maintenance to fit
into current OS FreeBSD releases,
L.S.Zolin and V.E.Kovtun, who had need for DAQ systems and allowed
to implement them using {\itshape ngdp},
S.G.Reznikov -- for cooperation during DAQ SPILL design and testing.

\vspace*{-5mm}

\renewcommand{\thesection}{A}
\renewcommand{\thesubsection}{A.\arabic{subsection}}
\section*{Appendix A.
}
\addcontentsline{toc}{section}{Appendix A.
}
\refstepcounter{section}
\label{ngdp.servnodes}

\vspace*{-3mm}

\subsection{{\itshape qdpb} inspired entities and imported elements}
\label{ngdp.servnodes.qdpb}

\vspace*{-2mm}

\hspace*{4mm} 
As we have already written in \cite{IsupJINRC10-34}, some {\itshape ngdp}
ideas are
inspired by {\itshape qdpb} design and some {\itshape qdpb} entities are
imported and redesigned for {\itshape ngdp}, as described below.\\
$\bullet$ The packet implementation changes after {\itshape qdpb}:\\
 -- the maximal packet size is enlarged up to 2048000~bytes;\\
 -- the control and answer packets are introduced by separate \verb|typedef|s
because
they should fit into single
Ethernet frame to guarantee the nonfragmented delivery;\\
 -- some functions (\verb|defrag_pack()|, \verb|merge_pack()| for kernel
context, \verb|recv_pack()| for user context) and macros (\verb|CHECK_ID()|,
\verb|CHECK_CRC()|, \verb|CHECK_PACK()| in two versions for both contexts,
\verb|CHECK_NUM()| for kernel context) are added to
{\bfseries\itshape packet(3)}%
, while some existing entities are refined;\\
 -- {\bfseries\itshape packet(3)} functions, originally implemented for the
32--bit {\ttfamily i386} (also known as {\ttfamily IA-32})\footnote{
Note, that to the {\ttfamily i386} \underline{architecture}
the \underline{CPU types} belong from {\ttfamily i386} itself through
{\ttfamily i486} and {\ttfamily i586} up to {\ttfamily i686}, which
includes Intel Pentium~4 and AMD K7.
} and {\ttfamily HPPA} architectures, are ported to {\ttfamily AMD64/EM64T}
architecture.\\
$\bullet$ The packet types support changes after {\itshape qdpb}:\\
 -- the packet type map as array of \verb|type_attr| structures
is introduced, that allows:\\
 -- the \verb|NEVTYPES| constant (number of packet types known
in the system) to be eliminated, because we always can count a number of
entries in the packet type map;\\
 -- the all values of the packet type attributes (currently -- the packet type
itself (\verb|u_short| value), the packet type name (C--string), the
offset in arrays (for backward compatibility), the packet number counter
(\verb|u_long| value), and the permissions to split the datafile before,
after or instead of this packet type writing) to be stored in the single
place at the map initialization instead of being spread over the conversion
functions of the packet type attributes; and\\
 -- this conversion functions for both contexts to be redesigned on the base
of the linear search in the map as implemented currently (see
{\itshape libsrc/pack\_types.c}),
or using fetch from the prefilled hash table of \verb|type_attr| structures
(see {\itshape test/thash.c} and section~\ref{ngdp.test})).

\vspace*{-5mm}

\subsection{Packet generator nodes {\bfseries\itshape ng\_mysource(4)}
and {\bfseries\itshape ng\_kthsource(4)}}
\label{ngdp.servnodes.mysrc}

\vspace*{-2mm}

\hspace*{4mm} The {\bfseries\itshape ng\_mysource(4)} node type is designed
very close to {\bfseries\itshape ng\_source(4)} from the standard
{\bfseries\itshape netgraph(4)} distribution, however it produces
{\itshape qdpb} / {\itshape ngdp} packets and is not coupled with the network
interface. The node generates packets by the handler executed by
{\bfseries\itshape callout(9)} mechanism (timer kernel thread activated at
each system clock interrupt), and accounts a size of the generated data in the
both packet (\verb|packets_out|) and byte (\verb|bytes_out|) units,
an elapsed astronomic time (\verb|elapsed|) between the moments of the generation
start and stop, a pure time (\verb|pure|) of the data generation and emission,
and a number of the occured transfer failures (\verb|fails|).

The {\bfseries\itshape ng\_mysource(4)} node
supports only one hook named \verb|output| simultaneously, however by default
persists after such hook disconnection, so it can be reused by \verb|connect|ing
\verb|output| again, and should be unloaded explicitly by \verb|shutdown|
generic control message.
The {\bfseries\itshape ng\_mysource(4)} can generate packets of only
one type simultaneously, because it supports only one value for \verb|type| and
for \verb|patt|. To overcome this limitation, the single hook named
\verb|input| is supported, and the data which arrived through it,
are sent untouched through the \verb|output| hook (if any).
This design allows to link a number of {\bfseries\itshape ng\_mysource(4)} nodes
with different configurations into a generation chain, which
is able to imitate the multityped data produced by some real source (f.e.,
the CAMAC interrupt handler).

The {\bfseries\itshape ng\_mysource(4)} understands the generic set of
control messages
and the following specific control messages as well:\\
\verb|start <int64_t num_of_packets>| -- starts the data generation;\\
\verb|stop| -- stops the data generation explicitly before the
  \verb|<num_of_packets>| generated;\\
\verb|getclrstats| -- returns the current statistics (values of
  \verb|packets_out|,
  \verb|bytes_out|, \verb|elapsed|, \verb|pure|, and \verb|fails|) and
  clears it;\\
\verb|getstats| / \verb|clrstats| -- returns / clears the current statistics
  (the same values);\\
\verb|setconf { packlen=<size> npack=<n> ticks=<m> pflag=<f> }| -- sets in
  according to the values of submitted \verb|struct ng_mysource_conf| members:\\
\hspace*{3mm} length \verb|packlen| of
  packets will be generated -- to \verb|<size>|,\\
\hspace*{3mm} number \verb|npack| of packets will be produced per one
  generator function call -- to \verb|<n>|,\\
\hspace*{3mm} time interval \verb|ticks| between two subsequent
  generator calls -- to \verb|<m>| ticks\footnote{
Tick is the time interval between two subsequent system clock interrupts,
usually 1/1000 sec.
  }, and\\
\hspace*{3mm} packet header flags\footnote{
After {\itshape qdpb} packet design the header flag field controls the valid
timestamp and CRC presence in the header. Possible flag values are
{\ttfamily \#define}d in {\itshape ng\_packet.h}~. Note, that if
{\ttfamily pflag} equal to zero is
supplied, the header flags are set to default ({\ttfamily F\_TIME | F\_CRC}),
while for {\ttfamily -1} -- to zero, that means to produce neither
timestamp nor CRC. All other values are bitwise ANDed with
{\ttfamily F\_MASK}.
} \verb|pflag| will be used for packets production -- to \verb|<f>|;\\
\verb|getconf| -- returns the current \verb|setconf| settings;\\
\verb|settype <uint16_t type>| -- sets the generated packets type to \verb|<type>|;\\
\verb|setpatt <uint64_t patt>| -- sets the filling pattern \verb|<patt>|
  for the packet bodies;\\
\verb|fragile <uint8_t flag>| -- (re)sets the boolean \verb|<flag>|, which
  requires to disconnect the \verb|input| hook during the \verb|output| one
  disconnection and to shutdown node without hooks, if true.
  This allows to easy \verb|shutdown| the whole generation chain of
  {\bfseries\itshape ng\_mysource(4)}s. However
  for freshly \verb|mkpeer|ed node this \verb|<flag>| by default is 0.

\verb|{get,clr,getclr}stats| control messages can be submitted and will be
processed during the data generation. 




The {\bfseries\itshape ng\_kthsource(4)} node type
preserves the described above functionality of the
{\bfseries\itshape ng\_mysource(4)} node type in general,
however, it uses the generic kernel thread {\bfseries\itshape kthread(9)}
mechanism instead of {\bfseries\itshape callout(9)}. This thread is very much
similar to a process in many aspects, however, it is not preempted by the
scheduling mechanism, so it generates packets as fast as possible.
If the packets transfer failure occurs, our thread voluntarily
participates in the scheduling: it {\bfseries\itshape msleep(9)}s during
\verb|<delay>| timeout\footnote{
Unfortunately, we have no means to {\bfseries\itshape wakeup(9)} it in
time, however a timeout value variation dependent of
failures rate or etc., can be implemented in principle.
} (by default $0.1$~sec.), and after that it continues the packets generation
and transfer.

The {\bfseries\itshape ng\_kthsource(4)} supports the same hooks as the
{\bfseries\itshape ng\_mysource(4)} and understands
the same control messages set. The only exception is
\verb|setconf|,
which submits \verb|struct ng_kthsource_conf|, where instead of \verb|ticks|
a \verb|delay| member is present. It means the
time interval (in ticks) for {\bfseries\itshape msleep(9)}ing
after the packet transfer failure.

\vspace*{-2mm}

\subsection{User context samples}
\label{ngdp.test}

\vspace*{-2mm}

\hspace*{4mm} To test and debug some aspects of
{\itshape ngdp} nodes in the user context, we implement a number
of program samples collected in the {\itshape test} directory:\\
{\itshape tbuf\_nNGO.c}
(queue disciplines for
{\bfseries\itshape ng\_fifo(4)} buffer implementation),\\
{\itshape thash.c} (implementation of hash table of \verb|struct type_attr|
and \verb|entrybyevtype()| function),\\
{\itshape tthr\_em.c} ({\bfseries\itshape ng\_em(4)} model with control
packets emission based on threads),\\
{\itshape tems.c} (``selfflow'' {\bfseries\itshape ng\_ems(4)} model
without thread),\\
{\itshape tthr\_p.c} ({\bfseries\itshape ng\_pool(4)} model with control
packets emission based on threads),\\
{\itshape tbp.c} (``selfflow'' {\bfseries\itshape ng\_bp(4)} model
without thread),\\
{\itshape twrap.c} (parser of some syntax suitable for configuration file),\\
\hspace*{4mm} as well as some user context program skeleton(s):\\
{\itshape tthr.c} (program with {\bfseries\itshape pthread(3)}),\\
{\itshape tmm.c} (example of {\bfseries\itshape ng\_mm(4)} using),\\
{\itshape tmmap.c} (parent/child data exchange through the mapped/shared memory
with se\-ma\-pho\-re synchronization),\ etc.

\vspace*{-5mm}


\end{document}